\documentclass[prb,preprint,amsmath,amssymb]{revtex4}
\def\dsp{\def\baselinestretch{1.2}\large\normalsize}
\usepackage{color}
\begin{document}
\author{Kevin Leung,$^*$ Anastasia G.~Ilgen, and Louise~J.~Criscenti}
\affiliation{Sandia National Laboratories, MS 1415, Albuquerque, NM 87185\\
$^*${\tt kleung@sandia.gov}}
\date{\today}
\title{Interplay of Physically Different Properties Leading to 
Challenges in Separating Lanthanide Cations
-- an {\it Ab Initio} Molecular Dynamics and Experimental Study}

\input epsf.sty
\dsp

 
\begin{abstract}

The lanthanide elements have well-documented similarities in their
chemical behavior, which makes the valuable trivalent lanthanide cations
(Ln$^{3+}$) particularly difficult to separate from each other in water.
In this work, we apply {\it ab initio} molecular dynamics
simulations to compare the free energies ($\Delta G_{\rm ads}$) associated
with the adsorption of lanthanide cations to silica surfaces at a pH condition
where SiO$^-$ groups are present.  The predicted $\Delta G_{\rm ads}$ for
lutetium (Lu$^{3+}$) and europium (Eu$^{3+}$) are similar within statistical
uncertainties; this is in qualitative agreement with our batch adsorption
measurements on silica.  This finding is remarkable because the two cations
exhibit hydration free energies ($\Delta G_{\rm hyd}$) that differ by $>$2~eV,
different hydration numbers, and different hydrolysis behavior far from silica
surfaces.  We observe that the similarity in Lu$^{3+}$ and Eu$^{3+}$
$\Delta G_{\rm ads}$ is the result of a delicate cancellation between the
difference in Eu$^{3+}$ and Lu$^{3+}$ hydration ($\Delta G_{\rm hyd}$), and
their difference in binding energies to silica.  We propose that disrupting
this cancellation at the two end points, either for adsorbed or completely
desorbed lanthanides (e.g., {\it via} nanoconfinment or mixed solvents), will
lead to effective Ln$^{3+}$ separation.

\end{abstract}
 
\maketitle
 
\section{Introduction}
\label{introduction}

Lanthanide (Ln) series elements exhibit specific utility in green energy
applications including lighting, wind turbines, electrified vehicles, and
catalysis.\cite{background} Improved mining and extraction techniques
are needed to expand the inventory of these critical elements.  Trivalent
lanthanide cations (Ln$^{3+}$) naturally occur as mixtures.  As such,
separation of Ln$^{3+}$ from each other is a technologically relevant and
chemically challenging problem.\cite{expt,expt1,lewis2015,qiang2016}
Ln separation schemes are complicated and hazardous, and have to be adjusted
depending on the composition of the Ln-containing ore.  Ion-exchange
separation of lanthanides has limited industrial use due to separation
efficiencies of existing resins not yielding sufficient Ln$^{3+}$
separation.\cite{book} Here we investigate what makes
adsorption more or less selective, so in the future we can develop
ion-exchange separation methods based on the fundamental science insight.

This separation problem has been addressed from several angles.
Extensive basic science research has been conducted to elucidate the solvation
properties of lanthanides in liquid
water.\cite{angelo2011,angelo2012,angelo2017,angelo2008,kuta2011,lnpka}
Further experimental\cite{expt,lewis2015,qiang2016} and theoretical
work\cite{comput1,comput2,comput3}
has focused on separating Ln$^{3+}$ from their mixtures using
organic ligands.  The results suggest that factors like pH, counter-ion
effects, soft-hard ion concepts, and ligand rigidity are 
important.\cite{expt,lewis2015,comput1,comput2,comput3,comput4}
Additional studies have focused on the possibility of using
silica and nanoporous silica materials (with and without functionalization
of their surfaces) to adsorb\cite{geiger2010,geiger2011,geckeis}
and separate\cite{silica1,giret} Ln$^{3+}$.  This latter pathway has proven
somewhat successful.  At pH$\sim$4, bare silica nanopores have been shown
to selectively adsorb
scandium,\cite{giret} but selectivity among lanthanides appears limited.
Ilgen has shown that, at pH$\sim$6, smaller lanthanides cations
are preferentially retained on nanoporous silica, depending on the nanopore
size.\cite{ilgen} At still higher pH values, lanthanides are less soluble
and precipitate as hydroxides.\cite{hydroxide}  Improvement of such
inorganic materials will likely enhance their Ln$^{3+}$ selectivity,
making them competitive with organic materials which are more costly,
less durable, and less environmentally benign.

Further studies are needed to determine how separation mechanisms determined
for organic ligands\cite{expt,lewis2015,comput1,comput2,comput3,comput4}
can be transferred to separation using inorganic materials.  Computationally,
molecular-level mechanistic studies have had difficulty determining
lanthanide bonding to inorganic surfaces or the relevant adsorption/desorption
free energies ($\Delta G_{\rm ads}$), in part because of a lack of classical
force fields that accurately reflect the interaction between $f$-electron
trivalent cations and water or silica surfaces.\cite{kuta2011}

In this work, we address the fundamental question of why different
Ln$^{3+}$ cations exhibit ``chemically similar'' adsorption equilibrium
constants on silica within a specific pH range, despite their significantly
different physical properties.  For example, lanthanides are known to exhibit
hydration free energies in liquid water ($\Delta G_{\rm hyd}$) that differ
significantly, by $>$2~eV.  Lutetium (Lu$^{3+}$) and europium (Eu$^{3+}$)
are at the endpoint
and midpoint of the $4f$-electron block of the periodic table, respectively.
Their ionic radii, differing by $\sim$0.1~\AA,\cite{angelo2011} are sufficient
to cause the Lu$^{3+}$ $\Delta G_{\rm hyd}$ to be more negative ({\it i.e.},
favorable) than that of Eu$^{3+}$ by
$\Delta \Delta G_{\rm hyd}$=-2.17~eV.\cite{comput4}   We will show that 
this difference far exceeds the relative adsorption free energy
($\Delta \Delta G_{\rm ads}$).  We will also demonstrate that the two
Ln$^{3+}$ exhibit different extent of hydrolysis reactions involving
splitting of H$_2$O molecules in their inner spheres; such concerted
hydrolysis/desorption behavior has been proposed and reported for other
multivalent cations.\cite{geiger2011,concerted1,concerted2,marmier1997,proton}
As a result, Ln$^{3+}$ hydration numbers ($N_{\rm hyd}$, the numbers of H$_2$O
molecules coordinated to different Ln$^{3+}$) can differ by two along parts of
the desorption reaction pathway.

We apply the {\it ab initio} molecular dynamics (AIMD) method, based on Density
Functional Theory (DFT), in conjunction with the potential-of-mean-force (PMF)
method which effectively extends AIMD time scales to yield accurate free
energies.  We show that Eu$^{3+}$ and Lu$^{3+}$ exhibit very similar 
$\Delta G_{\rm ads}$.  These predictions are supported by batch adsorption
measurements on non-porous silica surfaces.  The advantage of DFT is that it
can be further applied to analyze different energy contributions.
Our analysis leads us to conclude that Ln$^{3+}$
are not very similar; instead, the cancellation of large energy terms leads to
similar $\Delta G_{\rm ads}$ that hinders separation.  From this work, we
propose that the key to successful Ln$^{3+}$ separation technology resides in
the disruption of this cancellation in either the adsorbed or desorbed regime.

\color{black}
This work involves significant computational challenges.  The
accuracy of DFT functionals and pseudopotentials used to treat Ln$^{3+}$
needs to be addressed, and we will perform tests on one of the Ln$^{3+}$
pseudopotential used in this work.  However, there is an urgent need to
address the accuracy of other modeling details as well, such the
explicit treatment of both outer-shell solvating water molecules and
the environment near interfaces.  In implicit outer-shell
solvation DFT calculations,\cite{angelo2012,comput4} the predicted
$\Delta G_{\rm hyd}$ exhibit variations that are on the order of a fraction
of an eV.  The discrepancies partly arise from the use of different DFT
functionals or quantum chemistry methods, but variations in the implicit
solvation methods used also likely play a role.  In this work, we apply
exclusively explicit hydration treatment via AIMD.  We argue that such
statistical mechanically rigorous research needs to take place in parallel
with the development of more accurate DFT methods when dealing with Ln$^{3+}$
in aqueous interfacial environments.  AIMD calculations of $\Delta G_{\rm ads}$
at water/mineral interfaces for trivalent lanthanide cations, or for any
trivalent cations, remain rare and challenging because of the high charge
density involved.  Our work represents a key step in this direction.

\color{black}


\section{Method}

\subsection{Experimental Details}

Adsorption of Eu$^{3+}$ and Lu$^{3+}$ onto amorphous
silica was quantified. We used commercially available fumed silica (Sigma
Aldrich) with surface area of 192$\pm$3 m$^2$ g$^{-1}$ as reported in our
earlier work.\cite{knight2020} No background electrolyte or buffer were
used. Milli-Q water with the resistivity of 18 M$\Omega$$\cdot$cm
was used for all stock solutions and experiments. Lanthanide (Ln)
stock solutions were prepared by diluting their nitrate salts
Ln(NO$_3$)$_3$ in milli-Q water. Aqueous concentration was verified by
inductively coupled plasma mass spectrometry (ICP-MS) analysis as
described below. 

Initially 50$\pm$1 mg of silica was weighted into each centrifuge vial,
then 10 mL of Milli-Q H$_2$O was added, and samples were hydrated for
a minimum of 2 hours. To begin the adsorption experiment, lanthanide
stock solution was added, and the total volume of each sample was
brought to 50 mL. The pH was measured after 2 hours of reaction
and read at pH=5.0 $\pm$0.3 for all samples. Then the samples were
left on a shaker table for one-week (168 hours). Our preliminary
kinetics runs showed that adsorption equilibrium is reached after
48 hours. The initial concentration of each lanthanide in the
reactors was 0.1, 1, 10, 20, 30, and 50 $\mu$M L$^{-1}$. The sample with
10 $\mu$M L$^{-1}$ was made in triplicate to assess experimental error.
All experiments were performed at ambient temperature (22~$^o$C).

\subsection{AIMD Computational Details}
\label{aimd}

Finite temperature AIMD simulations apply the Perdew-Burke-Ernzerhof (PBE)
functional,\cite{pbe} the projector-augmented wave-based Vienna Atomic
Simulation Package (VASP),\cite{vasp1,vasp1a,vasp2,vasp3} a 400~eV energy
cutoff, and $\Gamma$-point sampling of the Brillouin zone.  A Nose thermostat
maintains the temperature at a slightly elevated 400~K.   A Born-Oppenheimer
energy convergence criterion of 10$^{-6}$~eV and a time step of 0.5~fs are
enforced.  These settings are similar to those in our previous ion desorption
work.\cite{concerted1}  
The charge-neutral simulation cell has a Si$_{40}$O$_{88}$H$_{13}^{3-}$
stoichiometry for the reconstructed $\beta$-cristobalite (001) slab, 123 H$_2$O
molecules, and a Lu$^{3+}$ or Eu$^{3+}$ cation in an initially bidentate
adsorbed configuration.  All simulation cells have
dimensions 14.32\AA$\times$14.32\AA$\times$26.0~\AA.  They represent a
$2^{1/2}$$\times$$2^{1/2}$ expansion of simulation cells we previously
applied.\cite{leung2009}  The larger simulation cells are adopted because
of the expectation that trivalent cations will experience stronger
image-image interactions at the same cell size.  
The VASP lanthanide pseudopotential used are ``Lu 23Dec2003'' 
and ``Eu 3 20Oct2008.''
  Lu$^{3+}$ has no unpaired $f$-electrons and the Eu
pseudopotential adopted (henceforth referred to as ``Eu(A)'') subsumes its $f$
electrons into the core; hence non-spin polarized DFT is applied for all AIMD 
simulations.  
\color{black} Calculations using the Lu pseudopotential, with explicit $4f$
electrons, are expected to be more accurate than those using the Eu(A)
pseudopotential, without explicit $4f$ electrons. \color{black}
Some static, spin-polarized DFT, DFT+U,\cite{dftu}
and HSE06\cite{hse06a,hse06b,hse06c} calculations, using the ``Eu 23Dec2003''
(``Eu(B)'') pseudopotential with an explicit, partially filled $f$-shell, are
conducted as spot checks (Sec.~\ref{pptest}).  

The number of H$_2$O molecules in the simulation cell is determined as follows.
Classical force field-based grand canonical Monte Carlo (GCMC)\cite{gcmc}
simulations are first applied to determine the average number of water
molecules filling the gap between the silica surfaces.\cite{leung2009}
The SPC/E water model,\cite{spce} a force field for silica based on
OPLS,\cite{pore_ff} and generic force field parameters pertinent to Ln$^{3+}$
are adopted for this purpose.  Silica atoms and the adsorbed cation are frozen
in DFT-optimized positions in GCMC calculations; only water molecules
are inserted into or removed from the simulation cell.  GCMC yields
7~H$_2$O molecules coordinated to the Ln$^{3+}$ adsorbed to the surface.
Switching to AIMD simulations and a Lu$^{3+}$ cation reduces this to four
after equilibration (Fig.~S2a).

The AIMD calculations in this work omit dispersion corrections.\cite{grimme}
This enables comparison with our previous pK$_a$ predictions which involve a
similar computational protocol.\cite{leung2009} Adding dispersion is known
to improve AIMD predictions of liquid water structure ($g(r)$) at
T=300~K.\cite{dftd_water1} But it has yet to be demonstrated that this
gives universally superior predictions at water/oxide interfaces.  

In the presence of acid functional groups at water/material interfaces, the
pH in the simulation cell should be pinned at the pK$_a$ of functional
groups, provided that (1) there is only one type of such groups; (2) a
fraction of them are deprotonated; (3) their pK$_a$ is lower than that
of H$_2$O; and (4) the surface groups do not interact with each other.
Within the non-interacting assumption, the pH in our AIMD cells should be
between 7.0~and 8.1 -- the pK$_a$ range previously predicted for this
single type of SiOH on this surface.\cite{leung2009} In experimental samples
with amorphous or crystalline silica, bimodal or trimodal pK$_a$ distributions
have been reported.\cite{gaigeot,shen,julie}  It would have been more
challenging to assign pK$_a$ in AIMD simulation cells with multiple types
of SiOH.

\subsection{Potential-of-Mean-Force Details, Reaction Coordinates}
\label{pmf_method}

The potential-of-mean-force profile is computed as
$\Delta W(Z)=-k_{\rm B}T \log P(Z)$ where $P(Z)$
is the probability that a $Z$ value is recorded in the trajectory within a
window, after making adjustments to rigorously remove the effect of umbrella
sampling penalties.  Here $Z$ is the coordinate normal to the silica-water
interface, $Z$=$z_{\rm Ln}-z_{\rm Si}$, Ln is the desorbing
lanthanide cation, and Si is the Si atom close to the two O$_{\rm Si}^-$
groups initially coordinated to the Ln$^{3+}$.  This coordinate is chosen to
accommodate the strong electrostatic attraction between trivalent cations and
surface silanol groups which can exhibit substantial bending motion.  
Harmonic penalties $A_o(Z-Z_i)^2$ are added to DFT energies in a series of
windows with a progression of $Z_i$ values, separated by 0.3~\AA\, spanning
the reaction paths. $A_o$ is set at 2~eV/\AA$^2$.

For Lu$^{3+}$, the initial configuration in the window has the
cation coordinated to two SiO$^-$ groups.  Then AIMD is applied.
Each subsequent window, with successively larger $Z_i$, and therefore
greater extent of desorption, is initiated by taking a 
configuration near the end of the trajectory from the previous window
along the $Z$-coordinate.  The first one or more picoseconds in each
window is used for equilibration only; statistics are collected for up to
45~ps.  Statistical uncertainties in $\Delta W(Z)$ are estimated
by splitting the trajectory in each window into five, calculating the
standard deviation in $\Delta W(Z)$ between the edge $Z$ values in each
window ($\Delta \Delta W(Z_i)$), and propagating the noise across windows
assuming gaussian statistics.  For Eu$^{3+}$ simulations, each sampling window
is initiated using an equilibrated configuration taken from the Lu$^{3+}$
trajectory with the same $Z_i$ value.  This generally entails an increase in
hydration number ($N_{\rm hyd}$) by one compared to Lu$^{3+}$, and sometimes
reduces the number of OH$^-$ groups coordinated to the Eu$^{3+}$ ($N_{\rm oh}$).
It typically takes 4-10~ps to equilibrate $N_{\rm hyd}$ for Eu$^{3+}$ in each
window.  Trajectory lengths in different umbrella sampling window are listed in
Table~S1 in the ESI.  The aggregate trajectory lengths used in all windows
exceed 336 and 255~ps for Lu$^{3+}$ and Eu$^{3+}$, respectively.

In the $Z_i$=5.0~\AA\, sampling window, the second of two O$_{\rm Si}$-Ln$^{3+}$
ionic bond is being broken, forming two valleys in the free energy landscape
with the absolute distance (not just $z$-coordinate) $R'$=$R_{\rm Ln-O}$
centered around $R'$=2.5~\AA\, (with one O$_{\rm Si}$-Ln$^{3+}$ bond)
and 4.2~\AA\, (with zero, Sec.~\ref{pmf_results}).  These
valleys are separated by a small free energy barrier.  In sampling
windows with $Z_i$$\ll$5.0~\AA, the $R'$$\sim$2.5~\AA\, valley is strongly
favored, while large $Z_i$ strongly favors the
$R'$$\sim$4.2\AA\, valley.  As our reaction coordinate $Z$ only constrains
the vertical distance between Ln$^{3+}$ and the designated Si atom, 
it does not yield a smooth transition between the two $R'$ valleys in
the handshake region near $R$$\sim$5.0~\AA\, (Sec.~\ref{pmf_results}).  

To deal with this problem and generate a smooth $\Delta W(Z)$, a secondary
umbrella sampling calculation is performed on reaction coordinate
$R'$= $R_{\rm Ln-O}$, as follows.  (a) By trial and error, we locate the
sampling window (or create a new sampling window) centered around $Z$=$Z_i$
where the $R_{\rm Ln-O}$$\sim$2.5~\AA\, and $R_{\rm Ln-O}$$\sim$4.0~\AA\,
valleys are similar in free energy.  This occurs at $Z_i$=5.0~\AA\, and
4.9~\AA\, for Eu and Lu, respectively.  (b) Keeping the primary umbrella
sampling $A_o$ and $Z_i$ parameters constnat, we introduce a series of harmonic
constraints $C_o$ ($R'$-$R_i$)$^2$, with $C_o$ chosen to be 1.5 or 
2.0 eV/\AA$^2$
and $R_i$ separated by between 0.2 to 0.4~\AA.  (c) Ideally, one would generate
a 2-dimensional PMF plot with $Z$ and $R'$.  In reality, the relatively
short AIMD simulations do not permit compiling accurate 2-D PMF statistics.
Instead, we align the $R'$ windows by integrating all $Z$ contributions
in a restricted range that feature in the two end-point $R'$ sampling windows,
so that there is overlapping statistics.  The ranges chosen are
4.73~\AA\,$<Z<$4.83~\AA\, for Lu$^{3+}$ and 4.87~\AA\,$<Z<$4.97~\AA\, for
Eu$^{3+}$.  (d) Integrating the $\Delta W(Z,R')$ over this narrow $Z$
range generates a pre-factor $F$ that describes the statistical
weight of the two $R'$ valleys.  (d) If $F$ is smaller than 1/15, we swtich to
larger $Z_i$ and retry (a)-(d).  If $F$ is larger than 15, we decrease $Z_i$
instead.  As mentioned above, $Z_i$=5.00~\AA\, and 4.90~\AA\, are chosen
for the Eu$^{3+}$ and Lu$^{3+}$ simulations via trial-and-error.  (e)
We add the probabilities $P(Z)$ from the two valleys, computed with $C_o$=0
(unconstrained in the $R'$ coordinate), weighted by the factor $F$.

A less severe version of this problem was encountered in our previous
calculations associated with Cu$^{2+}$ desorption.\cite{concerted1}
The Cu$^{2+}$-O(SiO$^-$) attraction is considerably weaker than the
Ln$^{3+}$-O(SiO$^-$) attraction, and a less elaborate procedure was 
devised to circumvent this issue.\cite{concerted1}  The higher local
charges associated with Ln$^{3+}$ cations makes AIMD PMF desorption 
calculations more challenging.

Unlike Ref.~\onlinecite{concerted1}, we do not use $Z=(z_{\rm O}-z_{\rm M})$
for reaction coordinate, where O is one of the O$_{\rm sio}^-$ groups initially
coordinated to the metal (``M'') cation.  The attraction between Ln$^{3+}$ and
the O atom in a surface SiO$^-$ group is stronger than in previous divalent
cation calculations.  As a result, using the previous
$Z=(z_{\rm O}-z_{\rm M})$ coordinate can lead to the O atom being pushed
into the silica interior while the Ln$^{3+}$ cation remains on the surface,
bonded to other surface silanol groups.  Unlike Ref.~\onlinecite{meijer2017},
we do not use the distance $R=|{\bf R}_{\rm O}-{\bf R}_{\rm M}|$.
This coordinate also allows the cation to roll along the surface
on to different surface sites, instead of away from the surface into the
bulk liquid. The water oxygen-cation coordination number
$N_{\rm hyd}$\cite{meijer2017} is not used for our purpose; it does not
distinguish possible outer-sphere (Ln$^{3+}$/H$_2$O/SiO$^-$) complexes, where
Ln$^{3+}$ and SiO$^-$ are 3.5-4.5~\AA\, apart, from Ln$^{3+}$
completely dissociated from SiO$^-$.\cite{criscenti2013}  Furthermore, for
Lu$^{3+}$, $N_{\rm hyd}$ does not increase monotonically as the cation desorbs.

Another, weaker harmonic potential of the form $B_o [(\delta x-x_o)^2 +
(\delta y-y_o)^2]$ constrains the Ln-O distances in the lateral directions.
Here $B_o$=0.025~eV, $\delta x$=$x_{\rm Ln}-x_{\rm O}$, $x_o$ is the
equilibrium value of $\delta x$ computed in completely unconstrained AIMD
trajectories, and $\delta y$ and $y_o$ are defined in analogous ways.

$\Delta W(Z)$ is effectively the constrained free energy at a $Z$ value; it
does not include the standard state reference associated with aqueous
solutions.  To obtain the adsorption free energy
($\Delta G_{\rm ads}$) from $\Delta W(Z)$, we integrate configuration space in
three dimensions, and account for  the entropic contribution from a standard
state 1.0~M ideal concentration solution:\cite{klein}
\begin{equation}
\Delta G_{\rm ads} /k_{\rm B} T = -\log \{  \int_\Omega d\Omega
                \exp [-\Delta W(Z)/k_{\rm B}T]  /(V_o)\} \, . \label{eq1}
\end{equation}
Here $V_o$ is the volume associated with 1.0~M aqueous solution (1662~\AA$^3$)
and T=300~K is assumed.  (``Standard state'' refers to [Ln$^{3+}$]=1.0~M;
no attempt is made to adjust the pH in Eq.~\ref{eq1}.)
The volume element $\Omega$ spans the configuration space where Ln$^{3+}$
is ``bonded'' to the SiO$^-$ group.  A limiting bonding distance of
3.20~\AA\, is assumed.  At this separation the pair correlation functions
between transition metal ions and water oxygen sites exhibit their first
correlation minima (Fig.~S1 in the ESI).  The angular distribution is also
involved in the
integral.  To our knowledge, $\Omega$ has not been standardized for PMF
calculations at interfaces.\cite{criscenti2013,meijer2017,kerisit2015} Here
we approximate it as a cylinder with a radius $R$=0.5~\AA.  Electrostatic
corrections associated with image dipoles are added to the PMF predictions by
creating a lattice model with screened coulomb interactions.\cite{concerted1}


We do not apply the metadyamics method, based on non-equilibrium
trajectories, to compute the PMF.\cite{meta} The umbrella sampling
approach used herein permits us to run trajectories of variable
lengths not determined ahead of time.  We examine statistical uncertainties
in each window to make sure there is no large, systematic drift in
$\Delta W(R)$ in each AIMD trajectory.  

\section{Results and Discussions}

\subsection{Batch Adsorption}

\begin{figure}
\centerline{\hbox{ \epsfxsize=3.20in \epsfbox{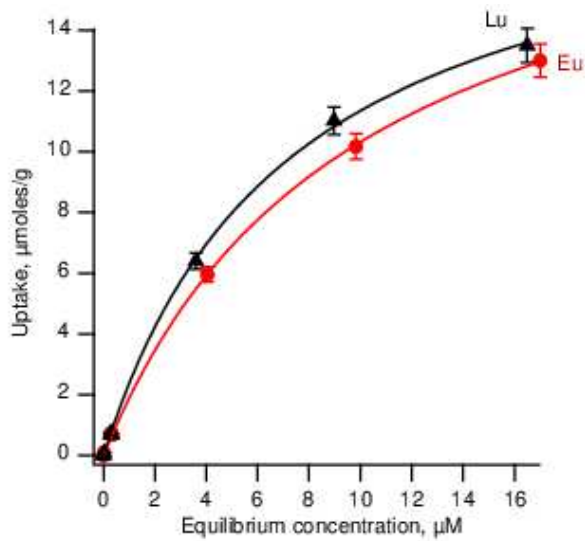} }}
\caption[]
{\label{fig1} \noindent
Uptake of Eu$^{3+}$ and Lu$^{3+}$ on amorphous silica SiO$_2$. 
Points = data, lines = fits for Langmuir isotherm equation.
Cumulative experimental error shown as error-bars was 4.2\%.
}
\end{figure}

First we discuss batch adsorption results on amorphous silica surfaces
that motivated this work (Fig.~\ref{fig1}).  The experiments were
performed on silica surfaces with $\sim$2 SiOH groups per nm$^2$ of 
surface, as discussed in our earlier publication.\cite{knight2020} 
The lanthanide adsorption data was fit using the Langmuir
isotherm model.  Based on the fitting, the maximum adsorption coverage
for Eu$^{3+}$ was estimated at 20.4~$\mu$moles/g, and the Langmuir constant
$K_{\rm L}$ was 0.102~L/g. For Lu$^{3+}$ the maximum adsorption was
estimated at 19.6 $\mu$moles/g, and $K_{\rm L}$ at 0.138~L/g.  The
overall affinity at pH~5.0 for Lu$^{3+}$ and Eu$^{3+}$ was similar, with
Lu$^{3+}$ being slightly more favorable.  This suggests that the experimental
$\Delta G_{\rm ads}$ for Eu$^{3+}$ and Lu$^{3+}$ are very similar.

\subsection{Potential-of-Mean-Force}
\label{pmf_results}

\begin{figure}
\centerline{\hbox{ (a) \epsfxsize=2.80in \epsfbox{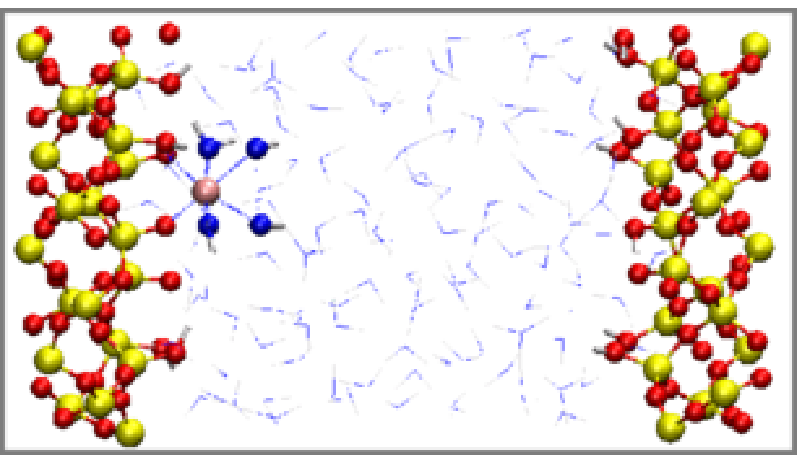} }}
\centerline{\hbox{ (b) \epsfxsize=1.40in \epsfbox{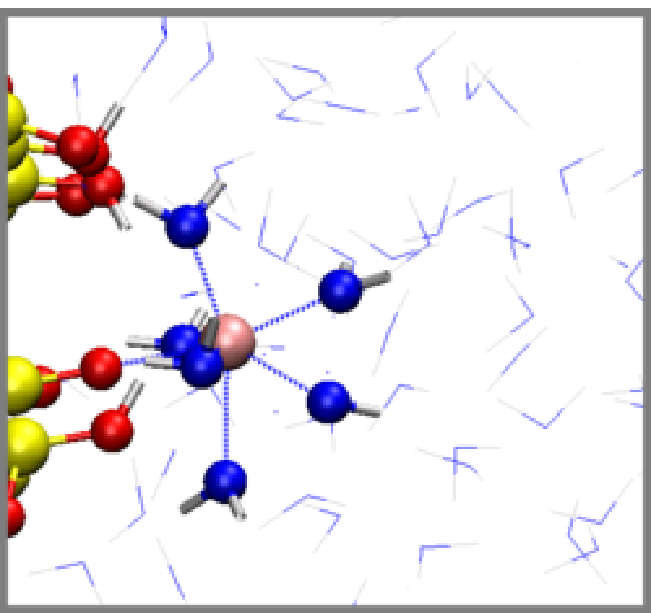} 
		   (c) \epsfxsize=1.40in \epsfbox{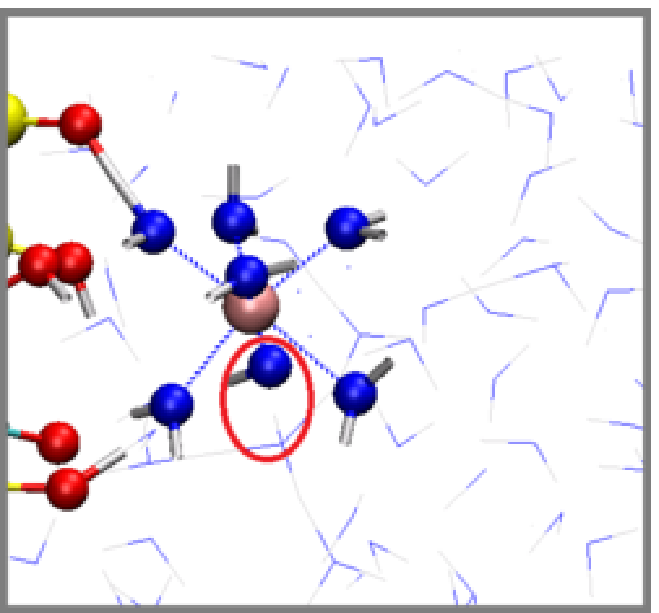} }}
\centerline{\hbox{ (d) \epsfxsize=1.40in \epsfbox{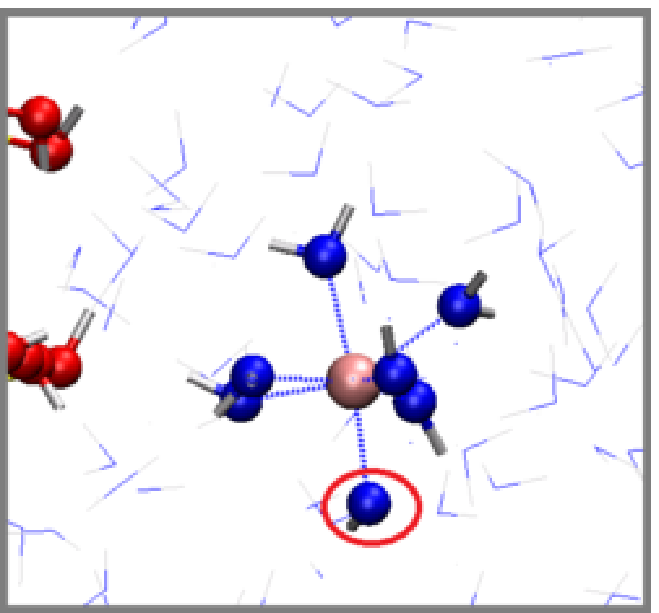} 
		   (e) \epsfxsize=1.40in \epsfbox{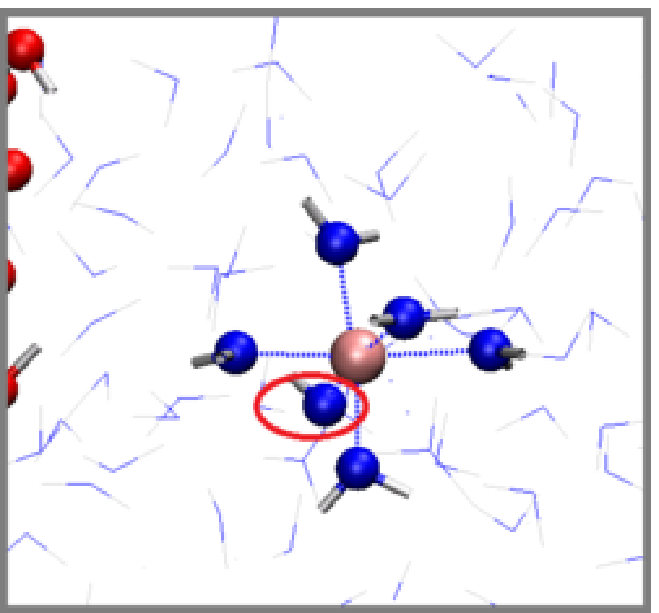} }}
\caption[]
{\label{fig2} \noindent
(a) The 14.3$\times$14.3$\times$26~\AA$^3$ simulation cell with adsorbed
Lu$^{3+}$ complex coordinated to two SiO$^-$ groups.  The reaction coordinate
$Z$ is along the $z$ direction (left-to-right).
(b)-(c) $Z$ centered at $Z_i$=4.9~\AA, with Lu$^{3+}$ bound to two or one
SiO$^-$ groups.  (d) $Z_i$=7.4~\AA.  (e) $Z_i$=8.0~\AA.  $Z_i$ is the center
of the constrained window.  Si, O, O (water), H, and Lu are depicted in
yellow, red, blue, white, and pink.  As some protons in H$_2$O are obscured,
the true OH$^-$ species are circled in red.  
}
\end{figure}

Next we turn to AIMD modeling.  Fig.~\ref{fig2}a depicts the charge-neutral
simulation cell containing 123~H$_2$O molecules and a partially deprotonated,
surface-reconstructed $\beta$-cristobalite slab; Lu$^{3+}$ is coordinated
to two SiO$^-$ groups on one surface.   Fig.~\ref{fig2}b-e depict Lu$^{3+}$
configurations in different PMF sampling windows as the cation desorbs.  The
SiOH surface density is 4.0/nm$^2$, and unlike in the experiments there are no
counter-ions.  Nevertheless, qualitative comparisons can be made.  

\begin{figure}
\centerline{\hbox{ \epsfxsize=3.50in \epsfbox{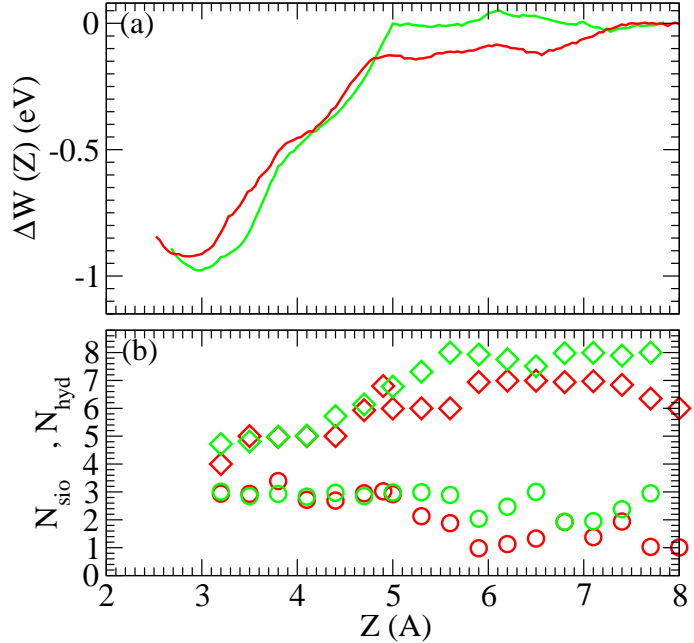} }}
\caption[]
{\label{fig3} \noindent
(a) Free energy profile along desorption reaction coordinate $Z$.
Red and green are for Lu$^{3+}$ and Eu$^{3+}$, respectively.
(b) Average number of SiO$^-$ groups ($N_{\rm sio}$, circles) in the
simulation cell, and the Ln$^{3+}$ hydration number ($N_{\rm hyd}$, diamonds).
The number of hydrolyzed H$_2$O molecules coordinated to the Ln$^{3+}$ is
$N_{\rm oh}$=($3$-$N_{\rm sio}$) on average.
}
\end{figure}

Fig.~\ref{fig3}a compares $\Delta W(Z)$ for Lu$^{3+}$ and Eu$^{3+}$.  The
shapes of $\Delta W(Z)$ at small $Z$ are similar, suggesting similar
energetics in the neighborhood of the optimal adsorption configuration.  
As desorption proceeds and $Z$ appoaches $Z$$\sim$5~\AA, a cross-over to a
quasi-plateau region is observed.  The Eu$^{3+}$ potential-of-mean-force
exhibits a slight repulsive behavior ($\Delta W(Z)$$>$0) near $Z$$=$5~\AA.
This is likely related to overscreening behavior associated with multivalent
electrolytes.\cite{overscreen} In contrast, due to hydrolysis in its hydration
shell (see below), the more weakly charged, hydrolyzed Lu$^{3+}$(OH$^-$)$_n$
complexes have a lower net charge and a monotonic $\Delta W(Z)$.  \color{black}
Our AIMD simulations are not ideally suited to investigating overscreening
effects due to the lack of counter-ions.  However, our future classical force
field MD simulations will reconsider possible overscreening.  \color{black}
Neither cation exhibits a local
minimum associated with outer sphere solvation; local minima may have
been helpful in engineering preferential adsorption motifs.

Integrating $\Delta W(Z)$ yields $\Delta G_{\rm ads}$=-0.79$\pm$0.04~eV
and -0.84$\pm$0.03~eV, assuming standard states for Lu$^{3+}$ and Eu$^{3+}$.
The uncertainties reflect one standard deviation.  Not included
in Fig.~\ref{fig3}a are electrostatic corrections.  As the Ln$^{3+}$ desorbs,
a significant dipole moment is created in the simulation cell, leading
to image-image repulsions in the lateral directions.
Using corrections based on lattice-models with dielectric
screening,\cite{concerted1} the magnitudes of $\Delta G_{\rm ads}$
are reduced by 0.01 and 0.09~eV for Lu$^{3+}$ and Eu$^{3+}$,
respectively.  These corrections change the preferred adsorption from
Eu$^{3+}$ over Lu$^{3+}$ to Lu$^{3+}$ over Eu$^{3+}$ by 0.03~eV, because
the Lu$^{3+}$(OH$^-$)$_n$ complex incurs less correction.  The computed
$\Delta G_{\rm ads}$ are larger in magnitude than our AIMD predictions of 
divalent metal cations on mineral surfaces, which range from -0.38 to
-0.71~eV.\cite{concerted1}  Our predicted Ln$^{3+}$ $\Delta G_{\rm ads}$ 
are more negative than those measured for some trivalent cations on
silica surfaces,\cite{geiger2011} likely because of the higher effective
pH in our simulations, our absence of counter-ions, and possible differences
in SiOH spatial distributions between our model and the experiment samples.
\color{black}
The measurements in Ref.~\onlinecite{geiger2011} are conducted at pH=4 while
the pH in our simulation cell is estimated to be $\sim$7.5.  A 3.5~unit
increase in pH translates into a maximum of
2$\times$3.5$\times$0.059~eV=0.41~eV increase in
Ln$^{3+}$ binding free energy.  This
estimate assumes that the increase in pH reduces the free energy needed to
deprotonate two neighboring SiOH groups, which then bind to the
Ln$^{3+}$.  It is much more difficult to quantify
the dependence of $\Delta G_{\rm ads}$ on surface structure details, such as
the distance between the two SiO$^-$ groups coordinated to the adsorbed
Ln$^{3+}$, without explicit AIMD/PMF simulations of the modified structure.
We plan to pursue AIMD simulations of monodentate Ln$^{3+}$ on model silica
surfaces in the future.  \color{black}

\begin{figure}
\centerline{\hbox{ \epsfxsize=4.00in \epsfbox{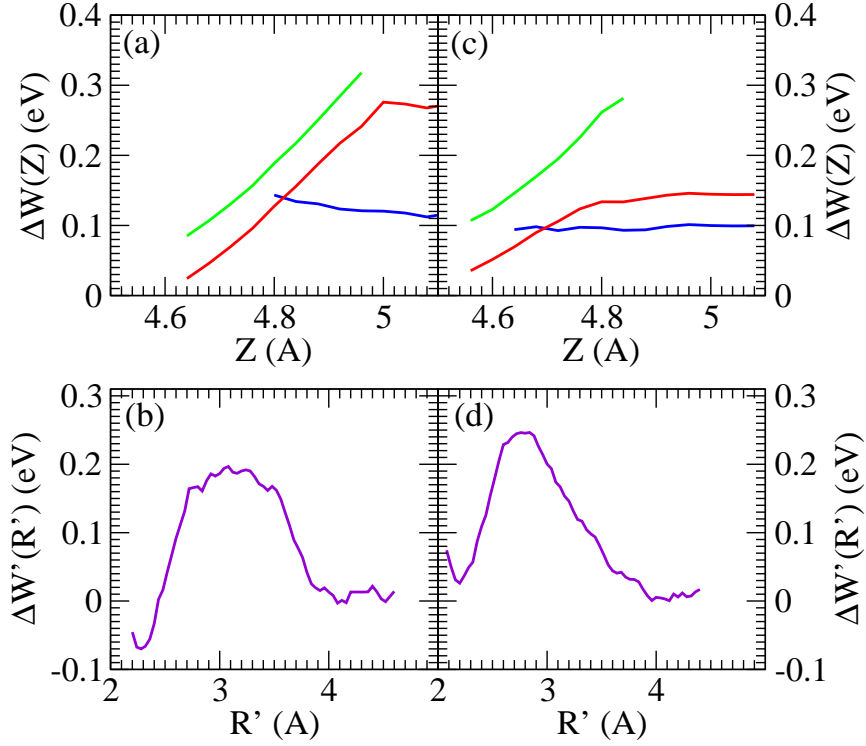} }}
\caption[]
{\label{fig4} \noindent
Panels (a)-(b) and (c)-(d) refer to $\Delta W(Z)$ for Eu$^{3+}$
and Lu$^{3+}$, with $Z_i$=5.0~\AA\, or $Z_i$=4.9~\AA, respectively.  
Green and blue are unnormalized $\Delta W(Z)$ segments in the two $R'$ free
energy valleys ith the same $Z_i$ constraint; they correspond to Ln$^{3+}$
coordinated to one or zero SiO$^-$.  Red depicts a weighted average of the
two, based on secondary umbrella sampling.  (b) and (d): potential-of-mean
force along a secondary reaction coordinate $R'$ ($\Delta W(R')$), with
$R'$ being one of the O-Ln distances. See text.
}
\end{figure}

As discussed in Sec.~\ref{pmf_method}, around $Z_i$=5.0~\AA, a secondary
PMF calculation, with another reaction coordinate $R'$, which is the
true distance between the Ln$^{3+}$ and a flagged O atom (not just its
$z$-component), is needed to augment our results.  Fig.~\ref{fig4}a shows
that the Eu$^{3+}$ $\Delta W(Z)$ in the $Z$$<$5.0~\AA\, and $Z$$>$5.0~\AA,
windows (green and blue lines) have different slopes.  Combining these
curves would yield a sharp kink in $\Delta W(Z)$.  A similar kink would have
occurred in the Lu$^{3+}$ $\Delta W(Z)$
(Fig.~\ref{fig4}c).  These kinks signal the inability of AIMD/PMF simulations
to reversibly sample two free energy valleys separated by small barriers.
Fig.~\ref{fig2}b-c show that these valleys are in fact associated with
Lu$^{3+}$ coordinated to one and zero SiO$^-$ groups, respectively; the local
minima are separated by Lu$^{3+}$ displacement parallel to the silica surface.
In terms of the secondary coordinate, $R'$ jumps from 2.5 to 4.2~\AA\,
between these valleys.  Our secondary PMF $\Delta W'(R')$
estimates the free energy differences between these two
valleys (Sec.~\ref{pmf_method}) and largely smooths over the kink in
$\Delta W(Z)$.  The barrier between the valleys are 0.2~to~0.25~eV
(Fig.~\ref{fig4}b,d), which are small but would have required much longer AIMD
trajectories to sample adequately without the secondary PMF.  While this 
approach involves approximations (Sec.~\ref{pmf_method}), the uncertainty 
in $\Delta \Delta G_{\rm ads}$ is lessened due to the expected cancellation
of errors between Lu$^{3+}$ and Eu$^{3+}$.

\subsection{Surprisingly Large Energy Difference in Adsorbed States}

The predicted preferential Lu$^{3+}$ adsorption is qualitatively consistent
with our batch adsorption measurements (Fig.~\ref{fig1}).  The difference
between the Lu$^{3+}$ and Eu$^{3+}$ $\Delta G_{\rm ads}$ is small, comparable
to the statistical uncertainty.  However, this small
$\Delta \Delta G_{\rm ads}$=-0.03~eV,
is surprising from an energetic standpoint -- despite the much-quoted
lanthanide ``chemical similarity.'' As mentioned above, Lu$^{3+}$ exhibits
$\Delta G_{\rm hyd}$ which is more favorable (negative) than the Eu$^{3+}$
value by -2.17~eV.\cite{comput4}  This represents the desorption end point
behavior.  For the two $\Delta G_{\rm ads}$ to be similar, there must be a
similarly large energetic difference at the adsorbed end point where Lu$^{3+}$
and Eu$^{3+}$ are in contact with silica.  In other words, chemical similarity
in fact derives from a cancellation of large (relative) energy terms.

\begin{figure}
\centerline{\hbox{ (a) \epsfxsize=2.00in \epsfbox{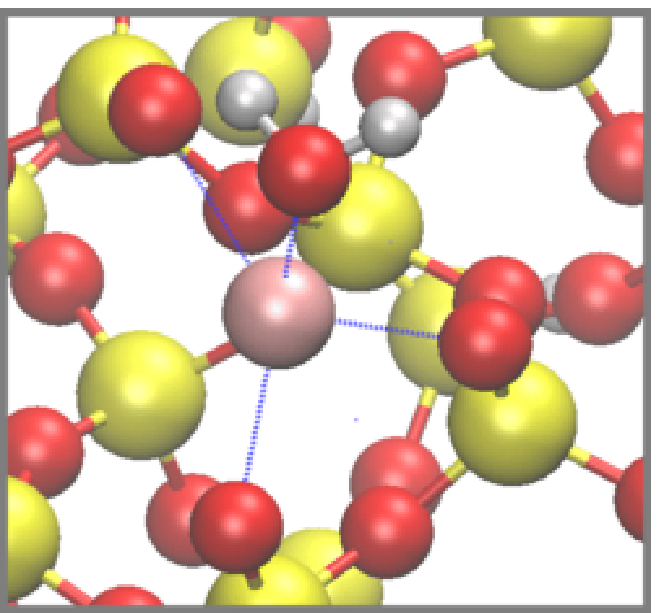} 
		   \epsfxsize=2.00in \epsfbox{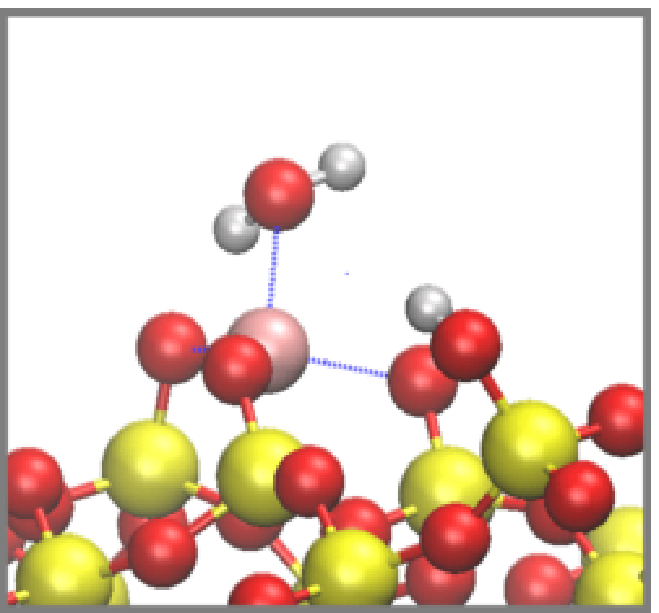} (b) }}
\centerline{\hbox{ (c) \epsfxsize=2.00in \epsfbox{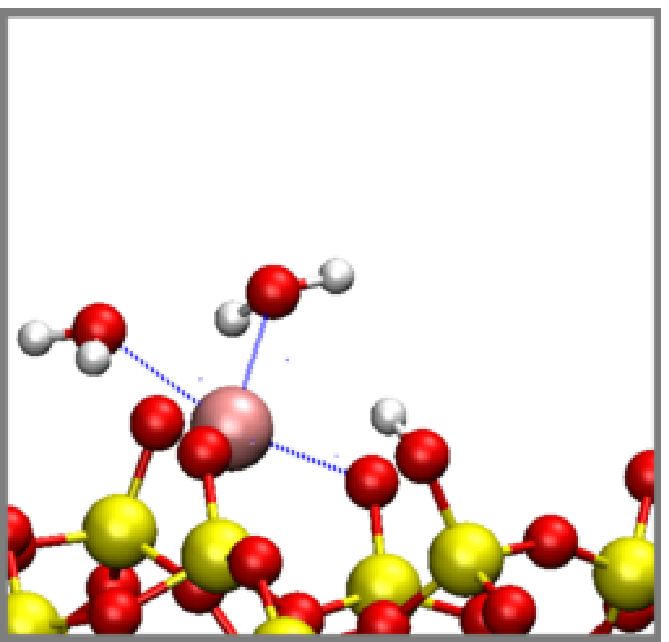} 
		   \epsfxsize=2.00in \epsfbox{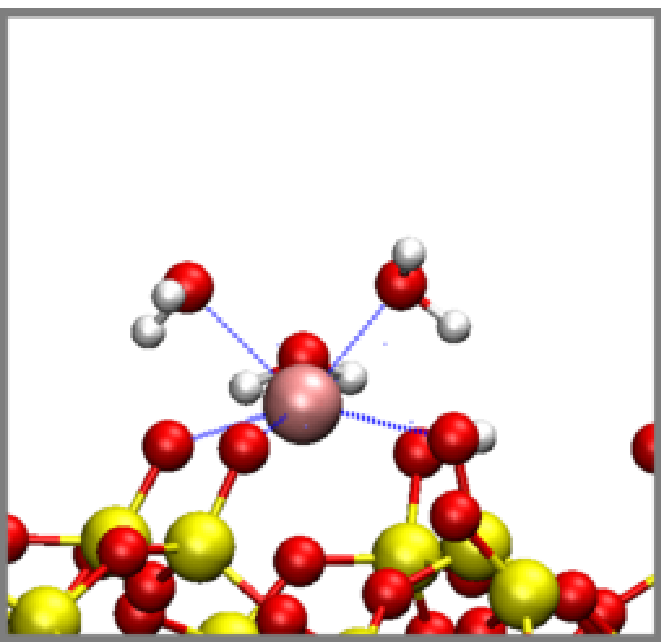} (d) }}
\caption[]
{\label{fig5} \noindent
(a)-(b) Top and side views of optimized Lu$^{3+}$ adsorption configuration
on silica surface, at T=0~K and coordinated to one H$_2$O molecule.
(c)-(d) Similar to (b) but with two and three H$_2$O coordinated to Eu$^{3+}$,
respectively.
}
\end{figure}

To examine this hypothesis, we approximate the energy difference 
($\Delta \Delta E_{\rm ads}$(dry)) in the adsorbed state by omitting most
water molecules.  We optimize configurations with a Lu$^{3+}$ or Eu$^{3+}$
cation at the binding site coordinated to two SiO$^-$ groups at T=0~K
(Fig.~\ref{fig5}a-b).  Only one H$_2$O molecule is included in the
simulation cell.  Maximally localized Wannier function analysis\cite{wannier}
confirms that, in these charge-neutral simulation cells with significant vacuum
regions, both lanthanides remain trivalent cations.  The net energy of the
Lu$^{3+}$ simulation cell is lower than that of Eu$^{3+}$ by 
$\Delta \Delta E_{\rm ads}$(dry)=-1.68~eV after substracting the
respective gas phase, bare ion energies.  This difference is indeed similar
to the reported -2.17~eV difference in $\Delta G_{\rm hyd}$.\cite{comput3}  
Note that we cannot report accurate absolute binding energies between
Ln$^{3+}$ and negatively charged silica because of the difficulty
in correcting the energies of slabs with net charges.

\color{black}
\subsection{Justification for Using the Eu(A) Pseudopotential}
\label{pptest}

The systems depicted in Fig.~\ref{fig5} also represent convenient
platforms to examine the validity of the Eu(A) pseudopotential used, which
omits $f$-electrons.  It is tempting to assume that Eu(B), which includes
$f$-electrons and requires the spin-polarized DFT method (Sec.~\ref{aimd}),
would be more accurate.  Our attempts at using Eu(B) in DFT/PBE-based
AIMD simulations, \color{black} however, result in occasional failures in
the self-consistent field procedure when performing convergence of the
Kohn-Sham densities and the Slater determinant orbitals. \color{black}
The likely reason is an unphysical charge transfer from silica to the Eu$^{3+}$
when using this pseudopotential during AIMD.  This is likely a failure of the
PBE functional, which is known to predict unphysical electron delocalization
in $f$-electron systems like CeO$_2$.\cite{anderson,nolan}  \color{black}
To deal with this problem, the DFT+U method\cite{dftu} has often been
applied as a remedy,\cite{lutfalla} as have hybrid DFT functionals like
HSE06.\cite{hse06a,hse06b,hse06c}

Here we compute the Eu(B)-predicted binding energies of a Eu$^{3+}$ relative
to Lu$^{3+}$ to the Si$_{40}$O$_{88}$H$_{13}^{3-}$ slab (Fig.~\ref{fig5}b) 
in vacuum, using the Eu(A) value as a reference.  If both Eu pseudopotentials
are equally accurate within the DFT/PBE framework, the energy difference
($\Delta \Delta E$) between them should be zero.  Instead, we find that the
Eu(B) result is favored by $\Delta \Delta E$=-0.70~eV over Eu(A).  When two
or three H$_2$O are included (Fig.~\ref{fig5}c-d), $\Delta \Delta E$=-0.51~eV
and -0.48~eV, respectively. 

Eu(B) gives consistently lower energies.  We argue that the significantly more
negative $\Delta \Delta E$ is consistent with unphysical hybridization between
silica and Eu(B) $4f$-orbitals, and/or possible electron transfer from silica
to Eu$^{3+}$.  Note that the extent of charge-transfer is non-trivial to
quantify; Wannier function analysis is challenging when using Eu(B), because
the partially-filled $f$ shell gives ``metallic'' behavior. To support our
argument, we turn to a rotationally invariant DFT+U approach\cite{dftu} with
$U$-$J$=4.5~eV.  We find that $\Delta \Delta E$=-0.13~eV, -0.11~eV, and -0.04~eV
with 1-3 H$_2$O in the simulation cell.  Therefore the Eu(A) pseudopotential
yields predictions very similar to Eu(B) which has $f$-electrons -- as long as
the more reliable\cite{lutfalla} DFT+U augmentation is applied to $f$-electrons
in the latter case.  Eu(A) does not have $4f$ electrons and DFT+U is
inapplicable there, while Lu has a full $4f$ shell and DFT+U is not expected
to yield results significantly different from PBE predictions.  We also note
that the Eu(A) and Eu(B) pseudopotentials have been shown to yield similar
structural properties when the latter is used in conjunction with DFT+U
augmentation.\cite{vogel}

Although applying DFT+U alongside the Eu(B) pseudopotential gives better
agreement with Eu(A), the results slightly vary with the value of ($U$-$J$).
Sec.~S3 in the SEI reports that changing ($U$-$J$) from 4.5~eV to 6.5~eV for
the one-water (Fig.~\ref{fig5}a-b) configuration changes $\Delta \Delta E$
from -0.13~eV to 0.05~eV.  Although this variation is only 0.18~eV, it is
significant compared with the $\Delta \Delta G_{\rm ads}$=0.03~eV
difference computed in AIMD simulations.  To lessen this ambiguity, we also
apply the HSE06 functional to both Lu$^{3+}$ and Eu$^{3+}$ bound to silica
surfaces.  The HSE06 $\Delta \Delta E$ for the Fig.~\ref{fig5}a-b configuration
is found to be -0.06~eV, which is close to zero.  We argue that the Eu(A)
pseudopotential, used in the majority of this paper, is an approximate way
to implement HSE06, or DFT+U with $U$-$J$=4.5~eV, on the Eu pseudopotential with
$f$-electrons.  It should in fact give more physical results than Eu(B) when
the latter is applied with PBE only.  From these state calculations, we
estimate that the systematic uncertainty associated with using the Eu(A)
pseudotpotential in our AIMD $\Delta \Delta G_{\rm ads}$ calculations is
between $\sim 0.06$ and 0.13~eV.
\color{black}  

\subsection{Hydrolysis and Hydration Behavior}

In the rest of this paper we analyze the differences in Lu$^{3+}$ and
Eu$^{3+}$ hydrolysis and hydration properties in an attempt to identify
factors that promote preferential Ln$^{3+}$ adsorption and separation.
As noted above, initially Lu$^{3+}$ is coordinated to two SiO$^-$
groups and 4~H$_2$O molecules (Fig.~\ref{fig2}a).  Bidentate adsorption is
consistent with dihydroxyl Yb$^{3+}$ coordination at low Yb$^{3+}$ coverage
known from previous analysis.\cite{sverjensky2008} 
In the cross-over region ($Z$$\sim$4.9~\AA,), there is an equilibrium between
two states, with Lu$^{3+}$ coordinated to one (Fig.~\ref{fig2}b) and zero
SiO$^-$ (Fig.~\ref{fig2}c) group. Hydrolysis only occurs above $Z$$\sim$5~\AA;
it arises from the transfer of proton(s) from H$_2$O molecule(s) coordinated
to the Lu$^{3+}$, to the silica surface {\it via} the Grotthuss mechanism.
When the desorption is almost complete (Fig.~\ref{fig2}d-e), Lu$^{3+}$ forms
a Lu$^{3+}$(H$_2$O)$_6$(OH$^-$), a Lu$^{3+}$(H$_2$O)$_5$(OH$^-$)$_2$, or
a Lu$^{3+}$(H$_2$O)$_4$(OH$^-$)$_2$ complex.

Eu$^{3+}$ configurations (not shown) are qualitatively similar.  Two main
differences are that Eu$^{3+}$ exhibits less tendency towards hydrolysis
than Lu$^{3+}$, and has larger hydration numbers ($N_{\rm hyd}$) throughout the
entire $Z$ range.  First we focus on hydrolysis.  Fig.~\ref{fig3}b reports
$N_{\rm sio}$ as $Z$ varies.  $N_{\rm sio}$ counts the number of deprotonated
Si-OH surface groups with a 1.25~\AA\, O-H cutoff distance in each sampling
window each centered at $Z$=$Z_i$.  $N_{\rm sio}$ plus $N_{\rm oh}$ (the
number of hydrolysis events  or number of OH$^-$ coordinated to Ln$^{3+}$)
should add to 3.0 on average.  
When $Z$$>$5~\AA, the average $N_{\rm sio}$ is larger in Eu$^{3+}$ simulation
cells compared to Lu$^{3+}$, meaning Eu$^{3+}$ induces less hydrolysis.  
Like Lu$^{3+}$, the onset in SiO$^-$ protonation state change is correlated
with the onset of the Eu$^{3+}$ $\Delta W(Z)$ plateau.

Based on the above analysis, it is surprising that a substantial cancellation
in energetics occurs for these two cations during desorption.  Although
the Lu$^{3+}$/SiO$_2$ (Fig.~\ref{fig5}a) and the Eu$^{3+}$/SiO$_2$ binding
configurations are similar (the latter is not shown), Lu$^{3+}$ is
coordinated to OH$^-$ while Eu$^{3+}$ is coordinated only to H$_2$O far from
the surface.  We conjecture that a substantial energy cancellation may arise
from the fact that the pK$_{a1}$ for hydrolysis is not far from the simulation
cell pH condition, so there is little change in $\Delta \Delta G_{\rm hyd}$ 
regardless of whether hydrolysis occurs.  Indeed, beyond $Z$=5~\AA, 
$\Delta W(Z)$ only changes by $\sim$0.1~eV in the $\Delta W(R)$ for both
cations.  Hence we argue that hydrolysis does not strongly affect the
predicted $\Delta G_{\rm ads}$ for either Eu$^{3+}$ or Lu$^{3+}$.  This is
unlike the case of Cu$^{2+}$.\cite{concerted1}
We propose that, if a {\it local} pH significantly higher than the
Ln$^{3+}$(H$_2$O)$_n$ pK$_{a1}$ can be maintained, this crucial
cancellation can be disrupted without causing Ln(OH)$_3$ precipitation,
leading to more selective Ln$^{3+}$ adsorption on silica surfaces.

Next we examine how differences in hydration numbers can affect
$\Delta \Delta G_{\rm ads}$.  Fig.~\ref{fig3}b also reports $N_{\rm hyd}$,
which is the number of H$_2$O or OH$^-$ oxygen atoms within 3.2~\AA\, of
Ln$^{3+}$.  It does not count coordination to SiO$^-$ groups.  Eu$^{3+}$
exhibits hydration numbers which exceed those of Lu$^{3+}$ by about 1-2
in the entire $Z$ range, i.e., $\Delta N_{\rm hyd}$$\sim$1-2. Note that
Ln$^{3+}$ hydration numbers in liquid water are measurable in X-ray and neutron
scattering experiments, and much computational effort has been devoted to
reproducing those values.  We do not observe Lu$^{3+}$(H$_2$O)$_8$ and
Eu$^{3+}$(H$_2$O)$_9$ complexes which have been
predicted in quantum chemistry or molecular dynamics
simulations.\cite{angelo2011,angelo2012,angelo2017,angelo2008} One reason is
the Lu$^{3+}$ inner sphere hydrolysis behavior discussed above, which has
not been accounted for in previous modeling work.\cite{comput4}  Indeed, 
metal cations coordinated to one or more OH$^-$ have been known to yield
reduced coordination numbers.\cite{concerted1,aloh4a}  \color{black}
Another reason may be that the explicit treatment of the second hydration
shell in AIMD simulations changes the first shell hydration number computed
using an implicit solvent approximation in the literature.
\color{black} Finally, different DFT functionals can yield $N_{\rm hyd}$ which 
are slightly different from each other and from experiments.\cite{jacs1}
The free energy difference associated with $N_{\rm hyd}$ values that differ
by one is generally on the order of k$_{\rm B}T$=0.025~eV, which
is small on our $\Delta W(Z)$ energy scale.  We propose that surface
constraints or functional groups that increase $\Delta N_{\rm hyd}$ to 
2~or higher may be needed to aid selective adsorption. 

\subsection{Sampling Dynamics}

\begin{figure}
\centerline{\hbox{ \epsfxsize=5.00in \epsfbox{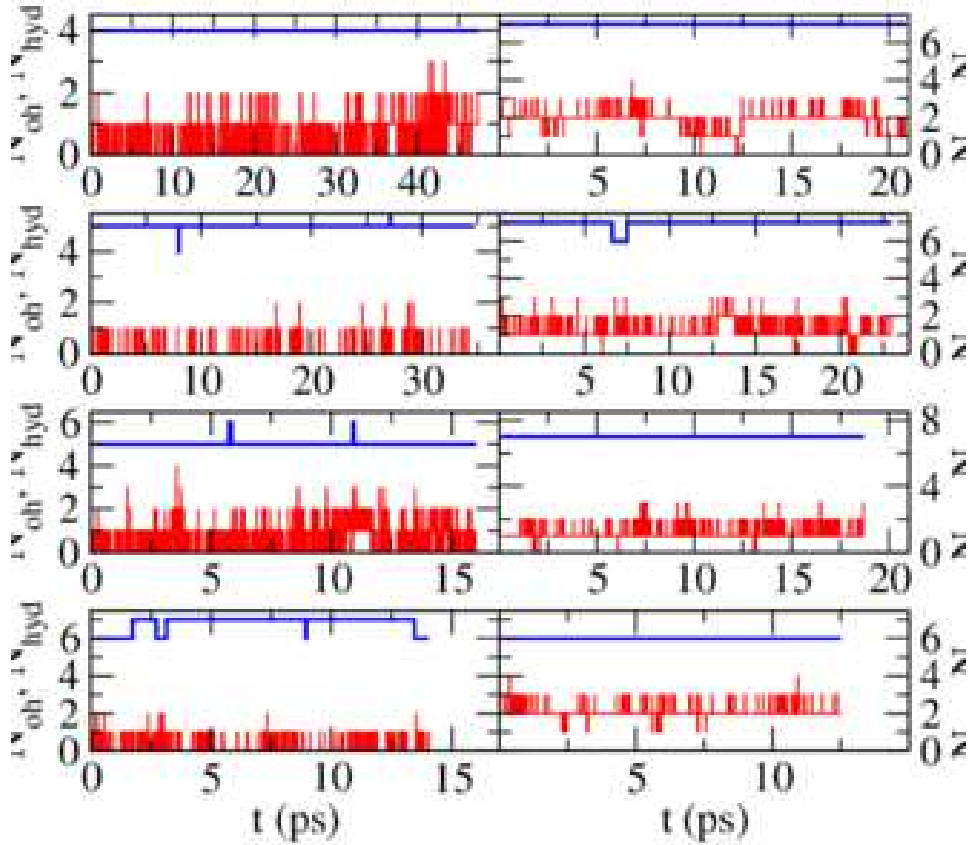} }}
\caption[]
{\label{fig6} \noindent
$N_{\rm oh}$ (red) and $N_{\rm hyd}$ (blue) as functions of time (ps) in
several Lu$^{3+}$ sampling windows.  From up to down, left column and
then right: unconstrained AIMD, $Z_i$=3.50~\AA, 3.80~\AA, 4.40~\AA,
5.60~\AA, 6.80~\AA, 7.40~\AA, and 8.00~\AA, respectively.
}
\end{figure}

For completeness, we briefly discuss dynamics.  $N_{\rm hyd}$ and $N_{\rm oh}$
for Lu$^{3+}$ are depicted as functions of time in selected sampling windows
in Fig.~\ref{fig6}.  (The complete set is given in the ESI.)  We have plotted
$N_{\rm oh}$ instead of $N_{\rm sio}$ here because, at certain times,
a H$^+$ may be in transit from the hydration shell of the desorbed
Ln$^{3+}$ to the silica surface; therefore $N_{\rm oh}(t)$ is more
descriptive than $N_{\rm sio}$ which equals (3-$N_{\rm oh}$) only on average.
At $Z$$>$5~\AA, the average $N_{\rm oh}$ is not monotonic as $Z$ varies, and
there are picosecond time scale proton exchanges between the Lu$^{3+}$ inner
hydration shell and the silica surface not accompanied by significant changes
in $\Delta W(Z)$ (Fig.~\ref{fig6}).  We never observe Si-OH groups from the
opposite surface of the silica slab being involved in acid-base reactions.  

Fluctuations of
$N_{\rm hyd}$ for trivalent cations can take long times at room temperature.
Fortunately, our sampling efficiency is improved by a slightly elevated
temperature of 400~K.  The fast proton transfer dynamics and the large
driving force guiding $N_{\rm hyd}$ evolution as $Z$ varies also help
inner sphere equilibration.  As a test, we have confirmed that, after
removing the umbrella sampling constraint for Lu$^{3+}$ at $Z_i$=3.5~\AA,
$N_{\rm hyd}$ spontaneously relaxes from 5 to its equilibrium value of
4 (Fig.~\ref{fig3}b) within 5~ps as the Lu$^{3+}$ relaxes to its adsorbed
configuration.  See also the ESI for discussions of the special case of the
$N_{\rm hyd}$ in one Lu$^{3+}$ window.  However, our simulation time scales
would have been insufficient to deal with the presence of anions, which
diffuse through water much more slowly than H$^+$.  This is the main reason
we have not included explicit counter anions in our AIMD simulations,
despite the fact that our batch adsorption experiments include NO$_3^-$.

\section{Conclusion}

In summary,
using AIMD/PMF calculations, we predict a slight, 0.03$\pm$0.05~eV preferential
adsorption of Lu$^{3+}$ over the lighter, larger Eu$^{3+}$ on model silica
surface with deprotonated silanol groups at its interface.  This finding is in
qualitative agreement with our batch adsorption work.  \color{black} The 0.03~eV
difference should be considered qualitative because of differences between
the experimental and computational systems, which include silica surface
details; pH; the presence/absence of NO$_3^-$ counter-ions; multiple surface
binding sites; and possible Ln$^{3+}$ dimerization.  We also note that
our Lu$^{3+}$ and Eu$^{3+}$ calculations utilize Lu and Eu pseudopotentials
with and without explicit $4f$-electrons, respectively.  As a result, the
Lu$^{3+}$ predictions are likely more accurate than the Eu$^{3+}$ predictions.
\color{black}
It is challenging to model Ln$^{3+}$ in aqueous media, both
because of DFT functional accuracy issues and because of the statstical
mechanical details that need to be implemented to obtain accurate free energy
changes associated with highly charged trivalent cations which can desorb
in conjunction with hydrolysis reactions.  In that sense, our pioneering AIMD
simulations pave the way for future examination of these details {\it via}
systematic variation of the model surface.  We explicitly address the role
of concerted proton motion, and give quantitative desorption free energy
predictions for trivalent lanthanide cation adsorption on mineral surfaces.  
\color{black} Our predictions provide guidance to molecular dynamics
simulations that apply classical force fields.  

AIMD simulations are first-principles in nature, and are generally more
accurate than force field-based MD.  They are computationally costly; the
necessarily limited trajectory lengths lead to unavoidable statistical
uncertainties.  A $\sim$10-fold preference for Lu$^{3+}$ over Eu$^{3+}$ at
T=300~K, measured under conditions slightly different from this
work,\cite{ilgen} translates into a -0.059~eV difference in
$\Delta G_{\rm ads}$, which is almost within AIMD noise level.  Instead of
resolving small $\Delta G_{\rm ads}$ differences, AIMD is most valuable at
providing mechanistic insights.  Thus it is more significant that the
similarity of the Lu$^{3+}$ and Eu$^{3+}$ desorption free energies is found
to arise from a cancellation between their relative adsorption energies
to deprotonated silica surface ($\Delta \Delta E_{\rm ads}$(dry)), and their
relative hydration free energies in liquid water ($\Delta \Delta G_{\rm hyd}$).
Although the cations exhibit substantial differences in hydration and hydrolysis
properties, the cancellation leads to very similar $\Delta G_{\rm ads}$.
To disrupt this cancellation of energy differences, enhance Ln$^{3+}$
selectivity, and aid separation, we propose that modification of the silica
surface to change the local pH or hydration environment, would be valuable.
Another approach suggested by our analysis is to modify $\Delta G_{\rm hyd}$
in the desorbed state, {\it e.g.}, by using a mixed solvent, or {\it via}
nanoconfinement.\cite{ilgen}

\section*{Conflicts of Interest}

Nothing to declare.

\section*{Acknowledgement}
 
We thank Jacob Harvey for his recommendation about Eu pseudopotentials and
other suggestions, and Jacquilyn Weeks for assistance with the manuscript.
This work is based on materials support by the U.S. DOE Office of Basic
Energy Sciences, Division of Chemical Sciences, Geosciences, and Biosciences
under Field Work Proposal Number 21-015452.
Sandia National Laboratories is a multi-mission laboratory managed and operated
by National Technology and Engineering Solutions of Sandia, LLC, a wholly owned
subsidiary of Honeywell International, Inc., for the U.S. Department of
Energy’s National Nuclear Security Administration under contract DE-NA0003525.
This paper describes objective technical results and analysis.  Any subjective
views or opinions that might be expressed in the document do not necessarily
represent the views of the U.S. Department of Energy or the United States
Government.

\newpage


\begin{references}

\bibitem{background}
J.H.L.~Voncken,  {\it The Rare Earth Elements: An Introduction.}
(Springer, 2015).



\bibitem{expt}
G.~Ferru, D.G.~Rodrigues, L.~Berthon, O.~Diat, P.~Bauduin and P.~Guilbaud,
{\it Angewandte Chem.~Int.~Ed.} {2014}, {\bf 53}, 5346-5350.

\bibitem{expt1}
I.~Lehman-Andino, J.~Su, K.E.~Papathanasiou, T.M.~Eaton, J.W.~Jian, D.~Dan,
T.E.~Albrecht-Schmitt, C.J.~Dares, E.R.~Batista, P.~Yang, J.K.~Gibson,
K.~Kavallieratos,
{\it Chem.~Commun.} 2019, {\bf 55}, 2441-2444.

\bibitem{lewis2015}
F.W.~Lewis, L.M.~Harwood, M.J.~Hudson, A.~Geist, V.N.~Kozhevnikov,
P.~Distler, J.~John,
{\it Chem. Sci.} 2015, {\bf 6}, 4812-4821.


\bibitem{qiang2016}
H.~Zhang, R.G.~McDowell, L.R.~Martin, Y.~Qiang,
{\it ACS Appl.~Mater.~Interfaces} 2016, {\it 8}, 9523-9531.

\bibitem{book}
N.~Krishnamurthy, C.K.~Gupta, {\it Extractive Metallurgy of
Rare Earths.} (2016, CRC press)

\bibitem{angelo2011}
P.~D'Angelo, A.~Zitolo, V.~Migliorati, G.~Chillemi, M.~Duvail, P.~Vitorge,
S.~Abadie and R.~Spezia,
{\it Inorg. Chem.} {2011}, {\bf 50}, 4572-4579.

\bibitem{angelo2012}
P.~D'Angelo and R.~Spezia,
{\it Eur. J. Chem.} {2012} {\bf 18}, 11162-11178.

\bibitem{angelo2017}
V.~Migliorati, A.~Serva, F.M.~Terenzio and P.~D'Angelo,
{\it Inorg. Chem.} {2017}, {\bf 56}, 6214-6224.

\bibitem{angelo2008}
I.~Persson, P.~D'Angelo, S.~De Panfilis, M.~Sandstr\"{o}m and L.~Eriksson,
{\it Chem. Eur. J.} {2008}, {\bf 14}, 3056-3066.

\bibitem{kuta2011}
J.~Kuta, M.C.F.~Wander, Z.~Wang, S.~Jiang, N.A.~Wall, A.E.~Clark, A.E.
{\it J.~Phys.~Chem.~C} {2011}, {\bf 115}, 21120-21127.



\bibitem{lnpka}
D.~Yu, R.~Du, J.-C.~Xiao, S.~Xu, C.~Rong and S.~Liu,
{\it J.~Phys.~Chem.~A} {2018}, {\bf 122}, 700-707.



\bibitem{comput1}
A.S.~Ivanov and V.S.~Bryantsev, {\it Eur.~J.~Inorg.~Chem.} {2016}, 3474-3479

\bibitem{comput2}
M.R.~Healy, A.S.~Ivanov, Y.~Karslyan, V.S.~Bryantsev, B.A.~Moyer and
S.~Jansone-Popova, {\it Chem. Eur. J.} {2019}, {\bf 25}, 6326-6331.

\bibitem{comput3}
B.~Sadhu and M.~Dolg, {\it Inorg. Chem.} {2019}, {\bf 58}, 9738-9748.

\bibitem{comput4}
J.~Ciupka, X.~Cao-Dolg, J.~Wiebke and M.~Dolg,
{\it Phys.~Chem.~Chem.~Phys.} {2010}, {\bf 12}, 13215-13223.


\bibitem{geiger2010}
D.S.~Jordan, J.N.~Malin and F.M.~Geiger,
{\it Environ.~Sci.~Technol.} {2010}, {\bf 44}, 5862-5867.

\bibitem{geiger2011}
D.S.~Jordan, S.A.~Saslow and F.M.~Geiger,
{\it J.~Phys.~Chem.~A} {2011}, {\bf 115}, 14438-14445.


\bibitem{geckeis}
D.~Garcia, J.~L\"{u}tzenkirchen, V.~Petrov, M.~Siebentritt,
D.~Schild, G.~Lefevre, T.~Rabung, M.~Altmaier, S.~Kalmykov,
L.~Duro and H.~Geckeis,
{\it Coll.~Surfaces~A} {2019}, {\bf 578}, 123610.

\bibitem{silica1}
Y.~Hu, E.~Drouin, D.~Lariviere, F.~Kleitz and F.-G.~Fontaine,
{\it ACS Appl. Mater. Interfaces} {2017}, {\bf 9}, 38584-38593.

\bibitem{giret}
S.~Giret, Y.~Hu, N.~Masoumifard, J.-F.~Boulanger, E.~Juere, F.~Kleitz and
D.~Lariviere, {\it ACS Appl. Mater. Interfaces}, {2018}, {\bf 10}, 448-457.

\bibitem{ilgen}
A.G.~Ilgen A.G. Non-provisional patent application
``Systems and Methods for Separating Rare Earth Elements Using
Mesoporous Materials.'' Filed on 3/11/2020.



\bibitem{hydroxide}
W.~Zhang and R.Q.~Honaker,
{\it Int.~J.~Coal~Geology} {2018}, {\bf 195}, 189-199.


\bibitem{marmier1997}
N.~Marmier, J.~Dumonceau and F.~Fromage,
{\it J.~Contaminant Hydrology} {\it 1997}, {\it 26}, 159-167.

\bibitem{concerted1}
K.~Leung, L.J.~Criscenti, A.W.~Knight, A.G.~Ilgen,
T.A.~Ho, and J.A.~Greathouse,
{\it J.~Phys.~Chem.~Lett.} {2018}, {\bf 9}, 5379-5385.

\bibitem{concerted2}
V.~Alexandrov and K.M.~Rosso,
{\it Phys.~Chem.~Chem.~Phys.} {2015}, {\bf 17}, 14518-14531.

\bibitem{proton}
Proton transfer-coupled processes, like redox reactions, have been prominent
in physical chemistry research, yielding unusual mechanisms that can be
exploited in catalysis.  See, e.g., 
J.~Cheng, X.~Liu, J.A.~Kattirtzi, J.~VondeVondele and M.~Sprik,
Aligning Electronic and Protonic Energy Levels of Proton-Coupled 
Electron Transfer in Water Oxidiation on Aqueous TiO$_2$.
{\it Angewandte Chem.~Int.~Ed.} {2014}, {\bf 53}, 12046-12050.

\bibitem{knight2020}
A.W.~Knight, P.~Ilani-Kashkouli, J.A.~Harvey, J.A.~Greathouse,
T.A.~Ho, N.~Kabengi and A.G.~Ilgen
{\it Environmental Science: Nano} {2020}, {\bf 7}, 68-80.

\bibitem{pbe}
J.P.~Perdew, K.~Burke, and M.~Ernzerhof,
{\it Phys. Rev. Lett.} {1996}, {\bf 77}, 3865-3868.

\bibitem{vasp1}
G.~Kresse, and J.~Furthm\"{u}ller,
{\it Phys.~Rev.~B} {1996}, {\bf 54}, 11169-11186.

\bibitem{vasp1a}
G.~Kresse, J.~Furthm\"{u}ller,~
{\it Comput.~Mater.~Sci.} {1996}, {\bf 6}, 15-50.

\bibitem{vasp2}
G.~Kresse and D.~Joubert,
{\it Phys.~Rev.~B} {1999}, {\bf 59}, 1758-1775.

\bibitem{vasp3}
J.~Paier, M.~Marsman and G.~Kresse,
{\it J. Chem. Phys.} {2007}, {\bf 127}, 024103.

\bibitem{leung2009}
K.~Leung, I.M.B.~Nielsen and L.J.~Criscenti,
{\it J. Am. Chem. Soc.}, {2009}, {\bf 131}, 18358-18365.

\bibitem{leung2013}
K.~Leung, L.J.~Criscenti,
{\it J.~Phys.~Condens.~Matter} {2012}, {\bf 24}, 124015.

\bibitem{dftu}
S.L.~Dudarev, G.A.~Botton, S.Y.~Savrasov, C.J.~Humphreys and A.P.~Sutton.
{\it Phys.~Rev.~B}, 1998, {\bf 57}, 1505.

\bibitem{hse06a}
Heyd, J.; Scuseria, G.E.; Ernzerhof, M.
Hybrid Functionals based on a Screened Coulomb Potential.
{\it J.~Chem.~Phys.} {\bf 2003}, {\it 118}, 8207-8215.

\bibitem{hse06b}
Heyd, J.; Scuseria, G.E.; Ernzerhof, M.
Hybrid Functionals Based on a Screened Coulomb Potential.
{\it J.~Chem.~Phys.} {\bf 2006}, {\it 124}, 219906.

\bibitem{hse06c}
Vydrov, O.A.; Heyd, J.; Krukau, A.V.; Scuseria, G.E.
Importance of Short-Range versus Long-Range Hartree Fock
Exchange for the Performance of Hybrid Density Functionals.
{\it J.~Chem.~Phys.}, {\bf 2006}, {\it 125}, 074106.

\bibitem{gcmc}
M.G.~Martin and A.P.~Thompson,
{\it Fluid Phase Equil.} {2004}, {\bf 217}, 105-110.

\bibitem{spce}
H.J.C.~Berendsen, J.R.~Grigera and T.P.~Straatsma,
{\it J.~Phys.~Chem.} {1987}, {\bf 91}, 6269-6271.

\bibitem{pore_ff}
K.~Leung, S.B.~Rempe, C.D.~Lorenz, {\it Phys. Rev. Lett.}
{2006}, {\bf 96}, 095504.

\bibitem{grimme}
See, e.g., S.~Grimme,
{\it J.~Comput.~Chem.}, {2006}, {\bf 27}, 1787-1799.

\bibitem{dftd_water1}
See, e.g., C.~Zhang, J.~Wu, G.~Galli and F.~Gygi,
{\it J. Chem. Theor. Comput.}, {2011}, {\bf 7}, 3054-3061.

\bibitem{gaigeot}
M.~Pfeiffer-Laplaud, M.P.~Gaigeot and M.~Sulpizi,
{\it J.~Phys~Chem.~Lett.} {2016}, {\bf 7}, 3229-3234.

\bibitem{shen}
Y.R.~Shen and V.~Ostroverkhov.
{\it Chem. Rev.} {2006}, {\bf 106}, 1140-1154.

\bibitem{julie}
A.M.~Darlington and J.~Gibbs-Davis
{\it J.~Phys.~Chem.~C} {2015}, {\bf 119}, 16560-16567.

\bibitem{meijer2017}
C.~Zhang, X.~Liu, X.~Lu, M.~He, E.J.~Meijer and R.~Wang,
{\it Geochim.~Cosmochim.~Acta} {2017}, {\bf 203}, 54-68.

\bibitem{criscenti2013}
L.E.~Katz, L.J.~Criscenti, C.-C.~Chen, J.P.~Larentzos and H.M.~Liljestrand,
{\it J.~Coll.~Interface Sci.} {2013}, {\bf 399}, 68-76.

\bibitem{klein}
J.~Blumberger and M.L.~Klein,
{\it Chem.~Phys.~Lett.} {2006}, {\bf 422}, 210-217.

\bibitem{kerisit2015}
S.~Kerisit, S.P.~Zarzycki and K.M.~Rosso,
{\it J.~Phys.~Chem.~C} {2015}, {\bf 119}, 9242-9252.

\bibitem{meta}
A.~Laio and M.~Parrinello, M.  
{\it Proc. Natl. Acad. Sci. USA} {2002}, {\bf 99}, 12562-12566.

\bibitem{overscreen}
N.R.~Haria and C.D.~Lorenz, 
{\it J.~Phys.~Chem.~C} {2015}, {\bf 119}, 12298-12311.

\bibitem{wannier}
N.~Marzari and D.~Vanderbilt, {\it Phys.~Rev.~B} {1997}, {\bf 56}, 12847.

\bibitem{anderson}
D.A.~Anderson, S.I.~Simak, B.~Johansson, I.A.~Abrikosov and N.V.~Skorodumova,
{\it Phys.~Rev. ~B} {2007}, {\bf 75}, 035109.

\bibitem{nolan}
M.~Nolan, S.~Grigoleit, D.C.~Sayle, S.C.~Parker and G.W.~Watson,
{\it Sur.~Sci.} {2005}, {\bf 576}, 217-229.

\bibitem{lutfalla}
S.~Lutfalla, V.~Shapovadov and A.T.~Bell, 
{\it J.~Chem.~Theoy Comput.} {2011}, {\bf 7}, 2218-2223.

\bibitem{vogel}
D.J.~Vogel, D.F.~Sava Gallis, T.M.~Nenoff and J.M.~Rimsza,
{\it Phys.~Chem.~Chem.~Phys.} {2019}, {\bf 21}, 23085.

\bibitem{sverjensky2008}
W.~Piasecki and D.A.~Sverjensky,
{\it Geochim.~Cosmochim.~Acta} {2008}, {\bf 72}, 3964-3979.


\bibitem{aloh4a}
T.R.~Graham, M.~Dembowski, E. Martinez-Baez, X.~Zhang, {\it et al.},
{\it Inorg.~Chem.} {2018}, {\bf 57}, 11864-11873.


\bibitem{jacs1}
K.~Leung and S.B.~Rempe, {\it J.~Am. Soc.~Chem.} {2004}, {\bf 126}, 344.




\end{references}
\end{document}